%% file: Chen-istc18-revised-arxiv.tex
\documentclass[conference,a4paper]{IEEEtran}

\usepackage{cite} 

 \ifCLASSOPTIONcompsoc
\usepackage[caption=false,font=normalsize,labelfont=sf,textfont=sf]{subfig}
 \else
\usepackage[caption=false,font=footnotesize]{subfig}
 \fi

\usepackage{amsmath,amssymb,amsfonts}
\usepackage{threeparttable} 
\usepackage{algorithmic} 
\usepackage{array} 
\usepackage[dvipdfmx]{xcolor} 
\usepackage{bm} 

\newtheorem{example}{Example}

\input{latexcommands}

\begin{document}
\title{Construction D$^\prime$ Lattices from Quasi-Cyclic Low-Density Parity-Check Codes}
       
%

%
\author{%
  \IEEEauthorblockN{Siyu Chen and Brian M. Kurkoski}
  \IEEEauthorblockA{Japan Advanced Institute of Science and Technology \\
                    1-1 Asahidai, Nomi, Ishikawa, Japan \\
                    Email: \{siyuch, kurkoski\}@jaist.ac.jp}
  \and
  \IEEEauthorblockN{Eirik Rosnes}
  \IEEEauthorblockA{Simula UiB, Bergen, Norway \\
                    Email: eirikrosnes@simula.no}
}

%


\maketitle

\IEEEpeerreviewmaketitle

\begin{abstract}
Recently, Branco da Silva and Silva described an efficient encoding and decoding algorithm for Construction D$^\prime$ lattices.
Using their algorithm, we propose a Construction D$^\prime$ lattice based on binary quasi-cyclic low-density parity-check (QC-LPDC) codes and single parity-check product codes. 
The underlying codes designed by the balanced-distances rule contribute in a balanced manner to the squared minimum distance of the constructed lattice, which results in a high lattice coding gain.
The proposed lattice based on IEEE 802.16e QC-LDPC codes is shown to provide competitive error-rate performance on the power-unconstrained additive white Gaussian noise channel.
\end{abstract}

%

\section{Introduction}\label{sec:INTRO}
Lattice codes are appealing for high data rate communications, 
because they can achieve both high coding gain and high shaping gain. 
Construction D$^\prime$ lattices \cite{Barnes-cjm83} are based on nested binary linear codes and can be regarded as multilevel codes \cite{Wachsmann-1999}.
If lattices are to be widely used in future communications systems, 
Construction D$^\prime$ lattices using binary low-density parity-check (LDPC) codes are an extremely likely candidate,
because many communication standards have recently adopted binary LDPC codes 
for error correction in the physical layer.
A particularly important class of LDPC codes are quasi-cyclic (QC) LDPC codes \cite{Fossorier-2004} that
have been included in various IEEE 802-related standards such as: 802.11n, 11ad, 15.3c, 16e and 22,
because of their excellent error-rate performance in noisy channels and efficient hardware implementation.

Recently, Branco da Silva and Silva solved an important problem, giving efficient encoding and decoding algorithms for Construction D$^\prime$ lattices using LDPC codes \cite{Silva-isit18}.
For lattice design, they used the \emph{equal-error-probability rule} to design binary LDPC codes by an extended progressive edge growth (PEG) algorithm that includes check-node splitting. 
Their multistage decoding algorithm has linear complexity and uses belief-propagation (BP) binary decoders.
Single-level Construction A lattices using QC-LDPC codes are proposed in \cite{Khodaiemehr-2017},
but since all the higher levels are uncoded, the error-rate performance quickly degrades as the block length increases.

In this paper, we use the \emph{balanced-distances rule} to design underlying codes to form Construction D$^\prime$ lattices.
The lattice coding gain can be maximized by 
designing the underlying codes with appropriate minimum Hamming distances.
Rather than randomly constructing binary LDPC codes by check-node splitting \cite{Silva-isit18},
we use binary QC-LDPC codes at the first encoding level and single parity-check (SPC) product codes at the second level.
One of our designs uses the QC-LDPC codes from the IEEE 802.16e standard \cite{802.16e-2006}. 
The minimum Hamming distance of these codes can be efficiently computed using their QC structure \cite{Rosnes-2009,Rosnes-2012}.

Compared to generalized Construction D$^\prime$ lattices \cite{Silva-isit18},
the proposed lattices are based on QC-LDPC codes without using the PEG algorithm. 
Our lattice design uses the minimum Hamming distance of binary codes and does not require the simulation-based design of the equal-error-probability rule.
QC-LDPC codes can be encoded with simple shift registers in linear time and efficiently decoded using partial parallelization.
SPC product codes are simple and easy to implement.
Simulation results show that the proposed lattice based on IEEE 802.16e QC-LDPC codes 
performs competitively to the generalized Construction D$^\prime$ lattice based on LDPC codes in \cite{Silva-isit18} and the Construction D lattice based on polar codes in \cite{Yan-isit13} over the power-unconstrained additive white Gaussian noise (AWGN) channel with respect to the block-error rate.

\section{Background} \label{sec:BG}

\subsection{QC-LDPC Codes}
An LDPC code is a linear block code given by 
the null space of a sparse $m \times n$ parity-check matrix $\H$.
If $\H$ consists of $z \times z$ square submatrices 
that are either circulant permutation matrices (CPMs) or the zero matrix,
the code can be characterized as a QC-LDPC code with circulant size $z$.
The $m \times n$ parity-check matrix of a QC-LDPC code with code length $n=zn_{\mathrm{b}}$ and redundancy $m=zm_\mathrm{b}$ can be represented by
\begin{align}
 \setlength{\arraycolsep}{2.05pt} 
 \H_{\mathrm{qc}} = \left[
  \begin{array}{cccc}
   \mathbf{P}_{b(0,0)} & \mathbf{P}_{b(0,1)} & \cdots & \mathbf{P}_{b(0,n_{\mathrm{b}}-1)} \\
   \mathbf{P}_{b(1,0)} & \mathbf{P}_{b(1,1)} & \cdots & \mathbf{P}_{b(1,n_{\mathrm{b}}-1)} \\
   \vdots              & \vdots              & \ddots & \vdots                  \\
   \mathbf{P}_{b(m_\mathrm{b}-1,0)} & \mathbf{P}_{b(m_\mathrm{b}-1,1)} & \cdots & \mathbf{P}_{b(m_\mathrm{b}-1,n_{\mathrm{b}}-1)}\\
  \end{array}
 \label{eq:H_qc}
\right],
 \setlength{\arraycolsep}{5pt} 
\end{align} 
where $\mathbf{P}_{b(i,j)}$, $0 \leq i \leq m_\mathrm{b}-1$, $0 \leq j \leq n_{\mathrm{b}}-1 $,
represents a $z \times z$ CPM,
which is a cyclic shift of the columns of the identity matrix to the right 
$b(i,j) \in \{-1,0,\ldots,z-1\}$ times; 
$\mathbf{P}_{-1}$ denotes the zero matrix.
In the following,
$\C_{\mathrm{qc}} = \left\{ \mathbf{c}_{\mathrm{qc}} \in \{0,1\}^n : \H_{\mathrm{qc}} \mathbf{c}_{\mathrm{qc}}^T = \bm{0} \right\}$,
where $(\cdot)^T$ denotes the transpose of its argument, 
will denote a binary QC-LDPC code with minimum Hamming distance $d_{\mathrm{qc}}$.
The code rate $r_{\mathrm{qc}}$ is lower-bounded by $1 - m/n$, 
with equality when $\H_{\mathrm{qc}}$ is full rank.

\subsection{SPC Product Codes}
A two-dimensional product code \cite[Sec.~3.5]{Ryan-Lin-book-2009} with code length $n = p q$ can be constructed from a $p \times q$ rectangular array in which 
every row and column are codewords of two binary linear block codes of
length $q$ and $p$, respectively.
Consider both component codes to be binary SPC codes, of which each codeword
consists of $q-1$ or $p-1$ information bits and one parity-check bit.
Then the product code formed from the SPC component codes is an SPC product code.
A codeword of the SPC product code consists of four parts:
$(p-1)(q-1)$ information bits, $p-1$ parity-check bits for the rows, 
$q-1$ parity-check bits for the columns and one parity-check bit for the row (or column) parity-check bits.
A $(p+q) \times pq$ parity-check matrix of an SPC product code is given by
\begin{align}
 \renewcommand{\arraystretch}{1.25} 
 \H_{\mathrm{spc}} = \left[
 \begin{array}{c}
  \bm{\mathcal{I}} \\
  \hline
  \bm{\mathcal{S}}
 \end{array}
 \right] = \left[
 \begin{array}{ccccc}
  \I_0         & \I_1         & \I_2         & \cdots & \I_{p-1}   \\
  \hline
  \mathbf{S}_0 & \mathbf{S}_1 & \mathbf{S}_2 & \cdots & \mathbf{S}_{p-1}
 \end{array}
 \right],
 \label{eq:H_spc}
\end{align}
where $\H_{\mathrm{spc}}$ is divided into two parts: $\bm{\mathcal{I}}$ and $\bm{\mathcal{S}}$
that each contains a single one in each column.
The top part $\bm{\mathcal{I}}$, 
which consists of identity matrices of size $q$, denoted by $\I_j$ ($0 \leq j \leq p-1$),
represents parity checks on the columns of the product code.
The bottom part $\bm{\mathcal{S}}$,
which consists of $p \times q$ submatrices denoted by $\mathbf{S}_j$ ($0 \leq j \leq p-1$), 
represents parity checks on the rows of the product code.
The submatrix $\mathbf{S}_j$ has $q$ contiguous ones in its $j$-th row and zeros elsewhere.   
This results in a ``staircase'' block row $\bm{\mathcal{S}}$ as follows:
\begin{align}
 \setlength{\arraycolsep}{2.25pt} 
 \bm{\mathcal{S}} \!=\! \left[
 \begin{array}{cccccccccccccccc}
  1 & 1 & \cdots & 1 & 0 &  0 & \cdots & 0 & 0 &  0 & \cdots & 0 & 0 & 0 & \cdots & 0 \\ 
  0 & 0 & \cdots & 0 & 1 &  1 & \cdots & 1 & 0 &  0 & \cdots & 0 & 0 & 0 &\cdots & 0 \\
  \vdots & \vdots & \vdots  & \vdots & \vdots & \vdots  & \vdots & \vdots & \ddots  & \ddots  & \ddots & \ddots & \vdots  & \vdots  & \vdots & \vdots \\ 
  0 & 0 & \cdots & 0 & 0 & 0 & \cdots & 0 & 0 &  0 & \cdots & 0 & 1 &  1 & \cdots & 1 \\ 
 \end{array}
 \label{eq:S}
 \right]\!.
 \setlength{\arraycolsep}{5pt}
\end{align}
In addition, the check on the checks of the SPC product code is contained in both $\bm{\mathcal{I}}$ and $\bm{\mathcal{S}}$, which leaves one redundant row in $\H_{\mathrm{spc}}$.
This results in $\H_{\mathrm{spc}}$ having rank $p+q-1$. 
In the following,
$\C_{\mathrm{spc}} = \left\{ \mathbf{c}_{\mathrm{spc}} \in \{0,1\}^n : \H_{\mathrm{spc}} \mathbf{c}_{\mathrm{spc}}^T = \bm{0} \right\}$
will denote a binary SPC product code with 
code rate $r_{\mathrm{spc}} = (p-1)(q-1) / pq$ and minimum Hamming distance $d_{\mathrm{spc}} = 4$. 

\subsection{Construction D$^\prime$ Lattices}
Construction D$^\prime$ converts a set of parity checks defining a family of 
nested binary linear codes into congruences for a lattice.
Let $\C_0 \subseteq \C_1 \subseteq \cdots \subseteq \C_{L} = \{0,1\}^n$ 
be a family of nested binary codes,
where $\C_l$ is an $(n, k_l, d_l)$ code of length $n$, dimension $k_l$ and minimum Hamming distance $d_l \geq 4^{L-l}$ for  $l = 0, \ldots, L-1$.
Let $\mathbf{h}_0, \ldots, \mathbf{h}_{n-1}$ be linearly independent vectors in $\{0,1\}^n$
such that for $l = 0, \ldots, L-1$, 
$\C_l$ with rate $r_l = k_l / n = (n-m_l) / n$ is defined by the parity-check matrix 
\begin{align}
 \renewcommand{\arraystretch}{1}
 \H_l = \left[
 \begin{array}{c}
  \mathbf{h}_0 \\
  \vdots \\
  \mathbf{h}_{m_l-1} \\
 \end{array}
 \right].
 \label{eq:H_l}
\end{align}
Note that because $\C_L = \{0,1\}^n$, $r_L = 1$, $k_L = n$, $m_L = 0$ and $d_L = 1$.
Then the Construction D$^\prime$ lattice $\Lambda$ is defined by 
\begin{align}
 \begin{split}
 \Lambda =& \left\{ \mathbf{x} \in \bbz^n : \mathbf{h}_j \cdot \mathbf{x}^T \equiv 0 \: (\mathrm{mod} \: 2^{l+1}),\right. \\
  &\left.\;\; 0 \leq l \leq L-1,\, m_{l+1} \leq j \leq m_l-1 \right\},
  \label{eq:congruence}
 \end{split}
\end{align}
where $\bbz$ denotes the set of integers. 
The volume of the Voronoi region for an $n$-dimensional 
Construction D$^\prime$ lattice $\Lambda$ is given by
\begin{align}
 V(\Lambda) &= 4^{\left(L- \sum^{L-1}_{l=0} r_l\right) n/2}.
\label{eq:volume}
\end{align}
The squared minimum Euclidean distance between any two points in the lattice $\Lambda$ is
the squared minimum distance $d^2_{\min}(\Lambda)$. 
Then the lattice coding gain is given by 
\begin{align}
\gamma_{\mathrm{c}}(\Lambda) = \frac{d^2_{\mathrm{min}}(\Lambda)} {V(\Lambda)^{2/n}}.
\label{eq:coding_gain}
\end{align}

\subsection{Design for Construction D$^\prime$ Lattices}
There are three approaches that have been used to design multilevel Construction D or D$^\prime$ lattices recently \cite{Wachsmann-1999}: 
1) The \emph{capacity rule} was used for designing the polar lattices \cite{Yan-isit13}:
       the component code rate $r_l$ is chosen equal to the capacity of 
       the channel at each coding level $l$, $l=0,\ldots,L-1$.
2) The \emph{equal-error-probability rule} was used for designing the LDPC lattice \cite{Silva-isit18}:
       the underlying binary codes should have an analytic expression for their error probability, 
       and the codes are chosen in such a way that the error probabilities of the channels or their bounds are equal.
3) The \emph{balanced-distances rule} is based on the squared minimum Euclidean distance in signal space.
       This rule is satisfied by well-known lattices such as the Barnes-Wall lattice.  
       

The balanced-distances rule used in this paper provides that
for Construction D$^\prime$ lattices at each encoding level $l$, 
the minimum Hamming distance $d_l$ of the component code $\C_l$ should satisfy
\begin{align}
 4^l d_l = \mathrm{constant}, \qquad l = 0, \ldots, L-1.
 \label{eq:BDR}
\end{align}
From the bound on the squared minimum distance of Construction D$^\prime$ lattices given in \cite[Th.~3.1]{Sadeghi-2006},
for a commonly used two-level construction, 
$d^2_{\min}(\Lambda)$ of the lattice $\Lambda$ given by Construction D$^\prime$ is bounded by  
\begin{align}
 4^L \geq d_{\min}^2(\Lambda) \geq \min \left\{ d_0, 4d_1 \right\},
 \label{eq:d_min_l=2}
\end{align}
from which $d^2_{\min}(\Lambda)$ is no greater than $16$ for $L=2$.
To achieve the upper bound of $d_{\min}^2(\Lambda) = 16$,
we set $\mathrm{constant}=16$ in \eqref{eq:BDR},
which results in $d_0 = 16$ for $\C_0$ and $d_1 = d_0 / 4 = 4$ for $\C_1$.
This will be our code design objective.

\subsection{Encoding and Decoding for Construction D$^\prime$ Lattices}
We use the encoding and decoding algorithms with complexity $O(Ln)$ for 
Construction D$^\prime$ lattices described in \cite{Silva-isit18}.
Sequential encoding modifies the congruence of Construction D$^\prime$ lattices 
from a zero vector to the syndrome vector of the previous encoding level.
Since the underlying component codes are binary LDPC codes,
one can easily modify the efficient encoding algorithm of \cite{Richardson-enco-2001} 
by appending the syndrome vector to the approximately triangular parity-check matrix of each component code at the left or right side.
Then a dummy bit `$1$' is correspondingly added to each component codeword at the head or the tail,
depending on the appended position of the syndrome vector.
On the other hand,
the decoding is a multistage decoding based on applying the sum-product algorithm (SPA) in each level.
Appending the syndrome vector to the parity-check matrix in each encoding level
makes no requirement of an efficient reencoding process for multistage decoding.
The log-likelihood ratio of the dummy bit `$1$' is set to infinity,
which indicates that the first or last bit of the component codeword is always equal to one.


\section{Proposed Construction D$^\prime$ Lattices}\label{sec:PP}
In this section, 
using the balanced-distances rule,
we propose two-level Construction D$^\prime$ lattices. 
The lattices are based on modified QC-LDPC codes $\C_{\mathrm{qc}}$ with $d_{\mathrm{qc}}=16$ for the first-level component code $\C_0$ and SPC product codes $\C_{\mathrm{spc}}$ with $d_{\mathrm{spc}}=4$ for the second-level component code $\C_1$.
These binary linear codes and their parity-check matrices are nested,
which means that $\C_0$ is contained in $\C_1$ as a subcode and $\H_1$ is a submatrix of $\H_0$. 
In contrast, 
a key property of \cite{Silva-isit18} is that $\H_1$ does not need to be a submatrix of $\H_0$.


\subsection{Construction}
First consider a QC-LDPC code $\C_{\mathrm{qc}}$ defined by the parity-check matrix $\H_{\mathrm{qc}}$ with circulant size $z$ as shown in \eqref{eq:H_qc}. 
$\C_{\mathrm{qc}}$ can be efficiently encoded by $\H_{\mathrm{qc}}$.
Then to construct a lattice, 
we want $\C_{\mathrm{qc}}$ to be nested with another binary linear code.
Our inspiration is from the fact that a code $\C_{\mathrm{spc}}$ can be naturally nested into $\C_{\mathrm{qc}}$
by merging their parity-check matrices.

For the proposed construction, $\H_{\mathrm{qc}}$ should contain at least one block row $i$ consisting of nonzero matrices, i.e., ${b(i,j)} \geq 0$ for all $0 \leq j \leq n_\mathrm{b}-1$.
The first-level code $\C_0$ is defined by a $(m + n / z) \times n$
parity-check matrix $\H_0$ of the following form:
\begin{align}
 \H_0 &= \left[
 \renewcommand{\arraystretch}{1.25} 
  \begin{array}{c}
  \H_{\mathrm{qc}} \\ 
  \hline
  \bm{\mathcal{S}} \\
 \end{array}
 \right] \nonumber \\
  &= \left[
  \setlength{\arraycolsep}{1.9pt} 
  \renewcommand{\arraystretch}{1} 
 \begin{array}{cccc}
  \mathbf{P}_{b(0,0)} & \mathbf{P}_{b(0,1)} & \cdots & \mathbf{P}_{b(0,n_{\mathrm{b}}-1)} \\
  \mathbf{P}_{b(1,0)} & \mathbf{P}_{b(1,1)} & \cdots & \mathbf{P}_{b(1,n_{\mathrm{b}}-1)} \\
  \vdots              & \vdots              & \ddots & \vdots                  \\
  \mathbf{P}_{b(m_\mathrm{b}-1,0)} & \mathbf{P}_{b(m_\mathrm{b}-1,1)} & \cdots & \mathbf{P}_{b(m_\mathrm{b}-1,n_{\mathrm{b}}-1)} \\
  \hline
  \mathbf{S}_{0} & \mathbf{S}_{1} & \cdots & \mathbf{S}_{n_{\mathrm{b}}-1} \\
 \end{array}
 \right],
 \label{eq:H_0}
 \setlength{\arraycolsep}{5pt} 
\end{align}
where the block row $\bm{\mathcal{S}}$ is as shown in \eqref{eq:S} with $p=n/z$ and $q=z$.
The resulting $\C_0$ is a binary linear code, 
specifically it is a modified QC-LDPC code $\C_{\mathrm{qc}}$ with 
code length $n$ and $m+n/z$ parity checks. 
Hence, the code rate of $\C_0$ is bounded as
$r_{\mathrm{qc}} \geq r_0 > 1 - (m+n/z) / n$,
where the left inequality is due to the appended $n/z$ rows in $\H_0$
and the right inequality indicates that $\H_0$ is \emph{rank deficient}.
Note that $\C_0$ still has a QC structure, 
since the parity check described by each $\mathbf S_j$ defines a cyclic code,
for all $0 \leq j \leq n_\mathrm{b}-1$.
Furthermore,
$d_0 \geq d_{\mathrm{qc}}$, since the appended $\bm{\mathcal S}$ cannot decrease the minimum Hamming distance. 
The second-level code $\C_1$, which contains $\C_0$, is defined by the $(z+n/z) \times n$ parity-check matrix
\begin{align}
 \setlength{\arraycolsep}{1.8pt}  
 \renewcommand{\arraystretch}{1.25} 
  \H_1 = \left[
 \begin{array}{c}
  \bm{\mathcal{P}} \\
  \hline
  \bm{\mathcal{S}}
 \end{array} 
  \right] {=} \left[
 \begin{array}{cccc}
  \mathbf{P}_{b(i,0)} & \mathbf{P}_{b(i,1)} & \cdots & \mathbf{P}_{b(i,n_{\mathrm{b}}-1)} \\
  \hline
  \mathbf{S}_{0} & \mathbf{S}_{1} & \cdots & \mathbf{S}_{n_{\mathrm{b}}-1} \\
 \end{array}
 \right],
 \label{eq:H_1}
 \setlength{\arraycolsep}{5pt}  
\end{align} 
which is a submatrix of $\H_0$.
Compared to \eqref{eq:H_spc}, 
although $\H_1$ does not necessarily include identity matrices,
$\C_1$ defined by $\H_1$ is still an SPC product code with code rate $r_1 = 1 - (z + n/z - 1) / n$.




\subsection{Design Example}
In this subsection,
we give a design example of the proposed Construction D$^\prime$ lattices.

\begin{example}
Consider a $(3,5)$-regular QC-LDPC code with circulant size $z = 34$, code length $n = 5 \times 34 = 170$ and $m = 3 \times 34 = 102$ parity checks. 
We start with a base-prototype matrix $\H_{\mathrm{b}}$ consisting of fifteen $34 \times 34$ identity matrices, i.e., $b(i,j) = 0$ for $0 \leq i \leq 2 $ and $0 \leq j \leq 4 $.  
We then replace the values of all $b(i,j)$ by random numbers from $\{0,\ldots,33\}$ and compute the corresponding minimum Hamming distance using the algorithm in \cite{Rosnes-2009,Rosnes-2012}, 
until $d_{\mathrm{qc}} = 16$ is obtained.
The $3 \times 5$ base-prototype matrix 
 \begin{align}
  \H_{\mathrm{b}} = \left[
  \begin{array}{ccccc}
   7&  13& 19&  22&  31 \\
   1&  11&  3&   2&  19 \\
   31&  25& 18&   3&  26 
  \end{array}
  \right],
  \label{eq:H_b}
 \end{align}
 which defines a code $\C_{\mathrm{qc}}$ with $d_{\mathrm{qc}} = 16$ for $z=34$, was obtained. 
Then we can obtain $\H_{\mathrm{qc}}$ by expanding $\H_{\mathrm{b}}$
and append $\bm{\mathcal{S}}$ to generate $\H_0$.
As a result,
we form a Construction D$^\prime$ lattice by two nested binary linear codes $\C_0$ and $\C_1$,
where $\C_0$ is a modified $\C_{\mathrm{qc}}$ with parameters $(170\!+\!1,68,d_0=16)$, $r_0 = 0.398$ and 
$\C_1$ is an SPC product code $\C_{\mathrm{spc}}$ with parameters $(170\!+\!1,132,d_1=4)$, $r_1 = 0.772$.
Thus from the balanced-distances rule, 
the constructed lattice $\Lambda$ has $d_{\min}^2(\Lambda) = 16$ and coding gain $\gamma_{\mathrm{c}}(\Lambda) = 7.04 \dB$.
 \label{exp:1}
\end{example}

Simulation results of each component code and the constructed lattice in Example~\ref{exp:1} are shown in Section~\ref{sec:SR}.

\section{Lattices from IEEE 802.16e QC-LDPC Codes} \label{sec:802.16e}


\subsection{IEEE 802.16e QC-LDPC Codes}
The IEEE 802.16e standard \cite{802.16e-2006} provides a class of well-designed QC-LDPC codes.
In the standard,
several $m_{\mathrm{b}} \times n_{\mathrm{b}}$ base-prototype matrices $\H_{\mathrm{b}}$ 
are used to generate QC-LDPC codes of various lengths and rates.
An $\H_{\mathrm{b}}$ is defined for the codes with length $n=2304$ and circulant size $z=96$. 
To generate codes of length $n$,
we can use the modulo operation
\begin{align}
 b_n(i,j) =  b_{2304}(i,j) \: \mathrm{mod} \: \left(96 \times \frac{n}{2304}\right)
 \label{eq:b_ij}
\end{align} 
to modify the right circulant permutations given by $b_n(i,j)$ that are specified as entries in $\H_{\mathrm{b}}$, for all $0 \leq i \leq m_\mathrm{b}-1$, $0 \leq j \leq n_{\mathrm{b}}-1$,
and circulant size $z = 96 \times n / 2304$.
From \cite{Rosnes-2012},
the rate-$1/2$ code of $n=1152$ and $z=48$ is a good candidate for our lattice construction, since $d_{\mathrm{qc}}=16$. 

\subsection{Modification to IEEE 802.16e QC-LDPC Codes}
We proposed to construct $\C_0$ by appending the block row $\bm{\mathcal{S}}$ to the parity-check matrix $\H_{\mathrm{qc}}$ 
in Section~\ref{sec:PP}.  
However, using the IEEE 802.16e base-prototype matrix,
it is not possible to form a single SPC product code for $\C_1$ using the appended $\bm{\mathcal{S}}$,
because the base-prototype matrix does not contain one block row of nonzero matrices.
Nevertheless, by modifying the base-prototype matrix it is possible to form two SPC-like product codes.
Then a concatenation of the two SPC-like product codes can be used for $\C_1$.
We intend to find block rows in the unmodified base-prototype matrix such that their sum can cover as many block columns as possible.
The selected block rows should have a minimal overlap of CPMs.

The modification to the $12 \times 24$ base-prototype matrix of rate-$1/2$ codes
is shown in Table~\ref{tb:H_b},
\begin{table*}[!htbp]
\centering
 \renewcommand{\arraystretch}{1.1} 
 \addtolength{\tabcolsep}{-2pt} 
 \caption{Base-Prototype Matrix $\H_{\mathrm{b}}$ of the First-Level Component Code $\C_0$ from IEEE 802.16e Rate-$1/2$ QC-LDPC Codes}
\label{tb:H_b}
\begin{tabular}{c|cccccccccccccccccccccccc}
$i \backslash j$ &0&1&2&3&4&5&6&7&8&9&10&11&12&13&14&15&16&17&18&19&20&21&22&23 \\
\hline 
0&-1&94&73&-1&-1&-1&-1&-1&55&83&-1&-1&7&0&-1&-1&-1&-1&-1&-1&-1&-1&-1&-1 \\
\bf{1}&-1&27&-1&-1&-1&22&$\mathbf{Q}_{-1}$&9&-1&-1&-1&12&$\mathbf{Q}_{33}$&0&0&-1&-1&-1&-1&-1&-1&-1&-1&-1 \\
2&-1&-1&-1&24&22&81&-1&33&-1&-1&-1&0&-1&-1&0&0&-1&-1&-1&-1&-1&-1&-1&-1 \\
3&61&-1&47&-1&-1&-1&-1&-1&65&25&-1&-1&-1&-1&-1&0&0&-1&-1&-1&-1&-1&-1&-1 \\
\bf{4}&-1&-1&39&-1&-1&-1&84&-1&-1&41&72&-1&-1&-1&-1&$\mathbf{Q}_{6}$&0&0&-1&-1&-1&-1&-1&-1 \\
5&-1&-1&-1&-1&46&40&-1&82&-1&-1&-1&79&0&-1&-1&-1&-1&0&0&-1&-1&-1&-1&-1 \\
6&-1&-1&95&53&-1&-1&-1&-1&-1&14&18&-1&-1&-1&-1&-1&-1&-1&0&0&-1&-1&-1&-1 \\
7&-1&11&73&-1&-1&-1&2&-1&-1&47&-1&-1&-1&-1&-1&-1&-1&-1&-1&0&0&-1&-1&-1 \\
\bf{8}&12&-1&-1&-1&83&24&-1&43&-1&-1&-1&51&-1&-1&-1&-1&-1&-1&$\mathbf{Q}_{10}$&-1&0&0&-1&-1 \\
9&-1&-1&-1&-1&-1&94&-1&59&-1&-1&70&72&-1&-1&-1&-1&-1&-1&-1&-1&-1&0&0&-1 \\
\bf{10}&-1&-1&7&65&-1&-1&-1&-1&39&49&-1&-1&-1&-1&-1&-1&-1&-1&-1&$\mathbf{Q}_{46}$&-1&-1&0&0 \\
11&43&-1&-1&-1&-1&66&-1&41&-1&-1&-1&26&7&-1&-1&-1&-1&-1&-1&-1&-1&-1&-1&0 \\
$\bm{\mathcal{S}}$&$\mathbf{S}_{0}$&$\mathbf{S}_{1}$&$\mathbf{S}_{2}$&$\mathbf{S}_{3}$&$\mathbf{S}_{4}$&$\mathbf{S}_{5}$&$\mathbf{S}_{6}$&$\mathbf{S}_{7}$&$\mathbf{S}_{8}$&$\mathbf{S}_{9}$&$\mathbf{S}_{10}$&$\mathbf{S}_{11}$&$\mathbf{S}_{12}$&$\mathbf{S}_{13}$&$\mathbf{S}_{14}$&$\mathbf{S}_{15}$&$\mathbf{S}_{16}$&$\mathbf{S}_{17}$&$\mathbf{S}_{18}$&$\mathbf{S}_{19}$&$\mathbf{S}_{20}$&$\mathbf{S}_{21}$&$\mathbf{S}_{22}$&$\mathbf{S}_{23}$\\ 
\end{tabular}
 \addtolength{\tabcolsep}{2pt} 
 \vspace{-0.2cm}
\end{table*}
where the appended block row $\bm{\mathcal{S}}$ consisting of $\mathbf{S}_j$, $0 \leq j \leq 23$, can be expanded to a $24 \times 1152$ submatrix of the form in \eqref{eq:S}.
We selected the block rows $1$, $4$, $8$ and $10$ to be modified by replacing four specific zero matrices (for which $b(i,j) = -1$)  
with the four CPMs:
$\mathbf{Q}_{b(1,12)}$, $\mathbf{Q}_{b(4,15)}$, $\mathbf{Q}_{b(8,18)}$ and $\mathbf{Q}_{b(10,19)}$.
To remove overlap of CPMs in one block column among the selected block rows,
a CPM $\mathbf{P}_{b(1,6)}$ is replaced with a zero matrix denoted by $\mathbf{Q}_{-1}$.
To ensure that $\H_{\mathrm{qc}}$ is free of cycles of length $4$,
we \emph{do not} reuse any existing or repeated value for these right circulant permutations.
For rate-$1/2$ codes with $n = 1152$, by random search, the 
selection $b_{n}(1,12)=33$, $b_{n}(4,15)=6$, $b_{n}(8,18) = 10$ and $b_{n}(10,19)=46$ for the four specific right circulant permutations mentioned above gave the lowest error rate for the modified $\C_{\rm qc}$. 
Furthermore,
by using the algorithm from \cite{Rosnes-2009,Rosnes-2012}, 
we found that the minimum Hamming distance of the modified $\C_{\mathrm{qc}}$ was increased to $d_{\mathrm{qc}}=23$.

\subsection{Lattice from Modified IEEE 802.16e QC-LDPC Codes}
For $\C_0$, we use the modified $\H_{\mathrm{b}}$ given in Table~\ref{tb:H_b} with $z=48$ resulting in a modified $\C_{\mathrm{qc}}$ with $n=1152$.
For $\C_1$, we add block rows $1$ and $8$ and block rows $4$ and $10$ of the modified $\H_{\mathrm{b}}$ and then
append the block row $\bm{\mathcal S}$.
The summations of block rows shown in Table~\ref{tb:H_c1} do not result in an SPC product code because of the double CPMs of weight $2$.
\begin{table*}[htbp]
 \centering
 \renewcommand{\arraystretch}{1.1} 
 \addtolength{\tabcolsep}{-3.2pt} 
 \caption{Base-Prototype Matrix of the Second-Level Component Code $\C_1$}
 \label{tb:H_c1}
 \begin{threeparttable}
  \begin{tabular}{c|cccccccccccccccccccccccc}
   &0&1&2&3&4&5&6&7&8&9&10&11&12&13&14&15&16&17&18&19&20&21&22&23 \\
   \hline
   $1+8$&12&27&-1&-1&83&$22/24$\tnote{a}&$\mathbf{Q}_{-1}$&$9/43$\tnote{a}&-1&-1&-1&$12/51$\tnote{a}&$\mathbf{Q}_{33}$&0&0&-1&-1&-1&$\mathbf{Q}_{10}$&-1&0&0&-1&-1\\
   $4+10$&-1&-1&$39/7$\tnote{a}&65&-1&-1&84&-1&39&$41/49$\tnote{a}&72&-1&-1&-1&-1&$\mathbf{Q}_{6}$&0&0&-1&$\mathbf{Q}_{46}$&-1&-1&0&0 \\
   $\bm{\mathcal{S}}$&$\mathbf{S}_{0}$&$\mathbf{S}_{1}$&$\mathbf{S}_{2}$&$\mathbf{S}_{3}$&$\mathbf{S}_{4}$&$\mathbf{S}_{5}$&$\mathbf{S}_{6}$&$\mathbf{S}_{7}$&$\mathbf{S}_{8}$&$\mathbf{S}_{9}$&$\mathbf{S}_{10}$&$\mathbf{S}_{11}$&$\mathbf{S}_{12}$&$\mathbf{S}_{13}$&$\mathbf{S}_{14}$&$\mathbf{S}_{15}$&$\mathbf{S}_{16}$&$\mathbf{S}_{17}$&$\mathbf{S}_{18}$&$\mathbf{S}_{19}$&$\mathbf{S}_{20}$&$\mathbf{S}_{21}$&$\mathbf{S}_{22}$&$\mathbf{S}_{23}$\\ 
  \end{tabular}
  \begin{tablenotes}
   \item [a] denotes a double CPM of weight $2$.
  \end{tablenotes}  
 \end{threeparttable} 
 \addtolength{\tabcolsep}{3.2pt} 
 \vspace{-0.25cm}
 \end{table*}
However, the resulting $\C_1$ is the concatenation of two SPC-like product codes, has $d_1 = 4$ and can be decoded using BP decoding. 
As a result,
the underlying code $\C_0$ is encoded with parameters $(1152\!+\!1, 564, d_0 \geq 25)$, $r_0 = 0.489$;
$\C_1$ is encoded with parameters $(1152\!+\!1, 1034, d_1=4)$, $r_1 = 0.897$.
For the constructed lattice $\Lambda$, 
we have $d_{\min}^2(\Lambda) = 16$ and $\gamma_{\mathrm{c}}(\Lambda) = 8.34 \dB$. 


\section{Simulation Results} \label{sec:SR}
To verify the contribution to the lattices of each component code 
described in Example~\ref{exp:1} and Section~\ref{sec:802.16e} without the effects of error propagation,
we evaluated the component codes separately from the lattice.
These component codes were used over an additive mod-$2$ Gaussian noise (AMGN) channel with noise variance $\sigma^2$.
Then we evaluated the error-rate performance of the lattices constructed by 
these nested component codes over the power-unconstrained AWGN channel.
In the simulation,
SPA decoding was performed in each stage for a maximum of $100$ iterations.
The block-error rate for each component code is a function of the signal-to-noise ratio defined as 
$\mathrm{SNR} = 1 / \sigma^2$.
The block-error rate for the lattice $\Lambda$ is a function of the volume-to-noise ratio defined as
\begin{align}
\mathrm{VNR} = \frac{V(\Lambda)^{2/n}}{2\pi {\mathrm e} \sigma^2}.
\label{eq:vnr}
\end{align}

\begin{figure}[tp]
\centering
\includegraphics[width=\columnwidth]{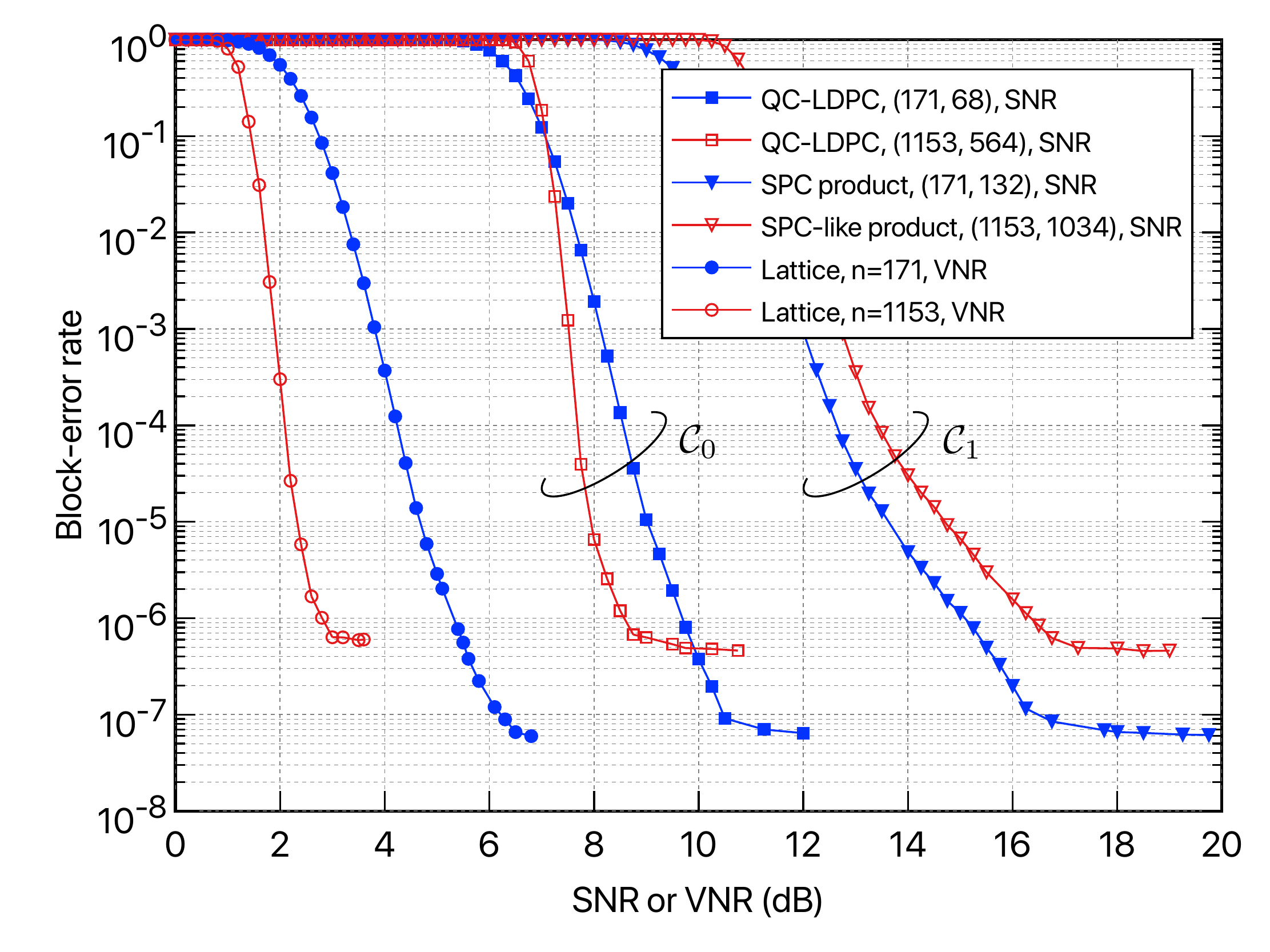}
\caption{Block-error rates of two sets of component codes and 
 the Construction D$^\prime$ lattices based on these codes.
 The $n=171$ set of codes are from Example~\ref{exp:1}. 
 The $n=1153$ set of codes are from a modified IEEE 802.16e QC-LDPC code.
 The component codes were individually used over the AMGN channel for different SNRs and 
 the lattices were used over the power-unconstrained AWGN channel for different VNRs.}
\label{fig:lattice-subcode-fer}
 \vspace{-0.05cm}
\end{figure}

Fig.~\ref{fig:lattice-subcode-fer} shows the block-error rates of 
two sets of component codes and the Construction D$^\prime$ lattices 
based on these binary codes.
For the $n=171$ set,
the error-rate performance of the constructed lattice is dominated by the first-level component code $\C_0$. 
This is so because the error-rate performance gap between the curves for $\C_0$ and $\C_1$ is always less than $6 \dB$, 
which is the difference in noise variance between the channels at the first and second encoding level.
For the $n=1153$ set,
the performance gap between the curves for $\C_0$ and $\C_1$ is $6 \dB$ at an error rate of $4 \times 10^{-5}$. 
Above this error rate, 
the performance of the constructed lattice is dominated by $\C_0$;
below this error rate,
performance is dominated by $\C_1$.

\begin{figure}[tbp]
 \centering
 \includegraphics[width=\columnwidth]{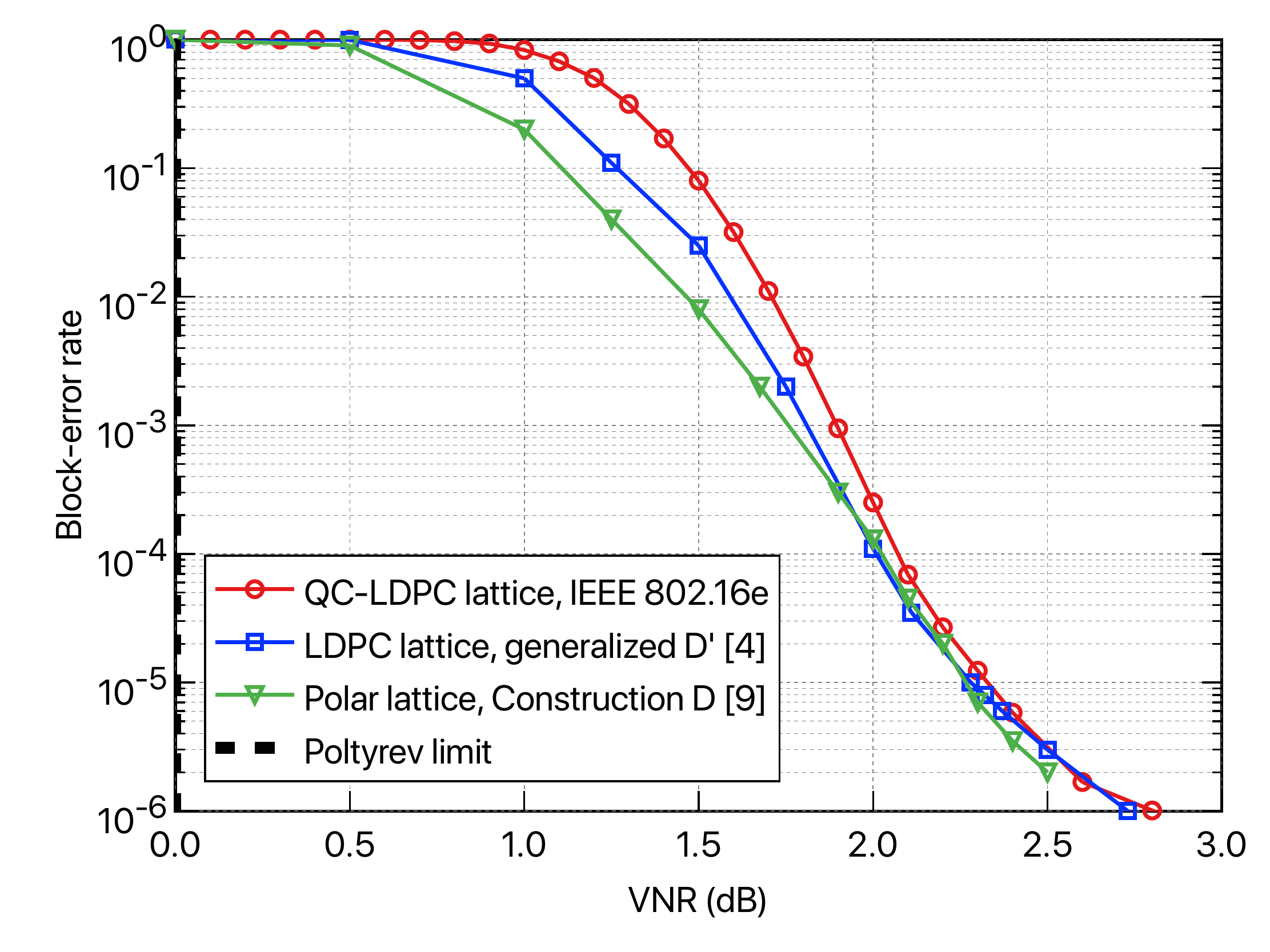}
 \caption{Block-error rates of two-level lattices
 used over the power-unconstrained AWGN channel with multistage decoding:
 the proposed lattice based on a modified IEEE 802.16e QC-LDPC code and a concatenation of two SPC-like product codes ($n = 1153$),
 a generalized Construction D$^\prime$ lattice based on LDPC codes ($n = 1025$, curve extracted from \cite{Silva-isit18}), 
 and a polar lattice ($n=1024$, curve extracted from \cite{Yan-isit13}).
 The Poltyrev limit which is at $0$ \textrm{dB} is also shown.}
 \label{fig:diff_lattice-fer}
  \vspace{-0.25cm}
\end{figure}

Then we compared the error-rate performance between the proposed lattice based on a modified IEEE 802.16e QC-LDPC code and a concatenation of two SPC-like product codes ($n=1153$), a generalized Construction D$^\prime$ lattice based on LDPC codes ($n=1025$; the simulation curve is extracted from \cite{Silva-isit18})
and a Construction D lattice based on polar codes ($n=1024$; the simulation curve is extracted from \cite{Yan-isit13}).
These two-level lattices were used over the power-unconstrained AWGN channel using multistage decoding.
Fig.~\ref{fig:diff_lattice-fer}
shows that the proposed lattice performs comparably to the generalized Construction D$^\prime$ lattice based on LDPC codes and looses about $0.1$ \textrm{dB} to the polar lattice.
The loss in performance might be due to the increase in $d_0$ that does not strictly satisfy the balanced-distances rule, which 
affected the coding gain $\gamma_{\mathrm c}(\Lambda)$ of the constructed lattice $\Lambda$.
However, decoding the proposed lattice has a low complexity $O(Ln)$ with sparse parity-check matrices;
decoding the polar lattice has complexity $O(Ln \log n)$.
The proposed lattice also benefits substantially from the QC structure of the underlying component codes.



\section{Conclusion} \label{sec:CC}
We have proposed a two-level Construction D$^\prime$ lattice using binary QC-LDPC codes and simple SPC product codes.
Lattices constructed by these component codes can benefit substantially from their QC structure.
The design criteria of the underlying component codes follow the balanced-distances rule.
Under this rule, the component codes contribute in a balanced manner to the squared minimum distance of the constructed lattices.
This results in a high coding gain for the lattices given by Construction D$^\prime$.
We modified a rate-$1/2$ QC-LDPC code from the IEEE 802.16e standard.
Simulation results show that the proposed two-level Construction D$^\prime$ lattice from the modified QC-LDPC code performs competitively to a generalized Construction D$^\prime$ lattice based on LDPC codes and a Construction D lattice based on polar codes in terms of block-error rate for the Poltyrev scenario.








\bibliographystyle{IEEEtran}
\bibliography{IEEEabrv,chen-bib,abbrev,bits}


\end{document}

%% file: latexcommands.tex
\usepackage{ifxetex}

\ifxetex

\else
   
   \newcommand{\C}{g}

\fi

\renewcommand{\C}{\mathcal C}

\usepackage{stmaryrd}

\usepackage[framed,numbered,autolinebreaks,useliterate]{matlab-prettifier}
\usepackage{listings}

\usepackage{etextools}
\usepackage{pgf,pgffor}
\def\rveccommas#1{ \big[ \foreach \x  [count=\ni] in {#1}  {%
\ifnum\ni=1%
\ \x
\else%
 , \x%
\fi%
} \ \big] }

\newlength{\outerwidth}
\newlength{\tbmargin}
\newlength{\bottommargin}

\newcommand{\keynote}[4]{
\setlength{\outerwidth}{#1}
\setlength{\tbmargin}{#2}
\setlength{\topmargin}{\tbmargin}
\setlength{\bottommargin}{\tbmargin}
\advance\topmargin by -24pt
\colorbox{white}{%
\fbox{\begin{minipage}{\outerwidth}
{\color{white} \rule{\linewidth}{0.4mm}\\[\topmargin]}
\begin{center}
\makebox[0pt][c]{
\advance\outerwidth by -#3
\begin{minipage}{\outerwidth}
\setlength\parindent{0cm}
\setlength\parskip{12pt}
#4
\vspace{\bottommargin}
\end{minipage}
}\end{center}%
\end{minipage}}
}
}

\newcommand{\bbz}{\mathbb Z}

\newcommand{\x}{\b x}

\newcommand{\I}{\mathbf I}

\renewcommand{\H}{\mathbf H}

\renewcommand{\b}[1]{\mathbf{#1}}

\newcommand{\dB}{\textrm{ dB}}

\usepackage{bm}

\definecolor{wesBlue}{HTML}{1DADE8}
\definecolor{wesRed}{HTML}{F34D29}